\documentstyle[aps,prl,twocolumn,epsf]{revtex}

\title{Composition Patterning in Systems Driven
by Competing Dynamics}
\author{Ra\'ul A. Enrique and Pascal Bellon}
\address{Department of Materials Science and Engineering and
Frederick Seitz Materials Research Laboratory, \\
University of Illinois at Urbana-Champaign, Urbana IL 61801}

\date{October 1999}

\begin{document}

\draft

\maketitle

\begin{abstract}
We study an alloy system where short-ranged, thermally-driven diffusion
competes with externally imposed, finite-ranged, athermal atomic exchanges,
as is the case in alloys under irradiation.
Using a Cahn-Hilliard-type approach, 
we show that when the range of these exchanges exceeds a critical value,
labyrinthine concentration patterns at a mesoscopic scale can be stabilized.
Furthermore, these steady-state patterns appear only for a window of the 
frequency of forced exchanges.  Our results suggest that ion beams 
may provide a novel route to stabilize and tune the size
of nanoscale structural features in materials.
\end{abstract}




The spontaneous formation of steady-state patterns have been extensively
observed in many equilibrium and nonequilibrium systems \cite{Cro93}.
While for equilibrium systems (e.g. ferrofluids, 
block copolymer melts, etc.) patterning is a result of the competition  
between repulsive and attractive interactions of different length scales,
in nonequilibrium systems (e.g. reaction-diffusion systems, etc.), 
steady-state patterning is often the result of the competition 
between several dynamical mechanisms.  A conceptual connection between 
the two classes of systems can sometimes be realized 
with the construction of Lyapunov functionals and effective Hamiltonians, 
by which steady-state pattern formation in dynamical systems
is interpreted as resulting from the competition between different types of 
{\it effective interactions}.

The kinetic Ising-type model with competing dynamics, and its continuum 
mean field counterpart, are instruments by which we hope to understand 
a whole class of nonequilibrium driven systems \cite{Mar98,Mar97}, 
ranging from fast ionic conductors to alloys under irradiation.  
The main ingredient of this model is the competition between 
two dynamics: one one hand, a thermally-driven mechanism trying to bring
the system to thermodynamical equilibrium; on the other hand, externally 
imposed particle exchanges of a nature essentially athermal.  
The usual attempt has been
to express the steady state of the system in terms of 
effective Hamiltonians and effective thermodynamic potentials.
P. Garrido, J. Marro, and collaborators \cite{Garr}, 
were able to derive effective Hamiltonians for several types of
1D Ising models with competing dynamics. 
Z. R{\'a}cz and collaborators \cite{Racz}, studied the relation between 
the range of the externally imposed exchanges and the range of
the effective interactions.  
In the context of alloys under irradiation, 
using a kinetic Ising-type model, Vaks and Kamyshenko \cite{Vak93} 
derived a formal expression for the steady state
probability distribution in terms of effective interactions, while from 
a continuum perspective, Martin \cite{Mar84} studied the corresponding 
dynamical phase diagram by an effective free energy. 
The possibility of patterning as a result of the 
competing dynamics has not been considered in these works. 
However, in the limiting case of arbitrary length external exchanges,
patterning has been recently observed in mean field and Monte Carlo
simulations \cite{Pav97}. 
In this situation, the
coarsening of segregated phases (magnetic domains in the Ising case)
saturates, leading to a steady-state labyrinthine patterning 
at a mesoscopic length scale. This microstructure is rationalized in
terms of a competition between the attractive nearest neighbors 
interactions, and a repulsive electrostatic-like effective interaction.
These patterns do not appear if the external exchanges are 
short range, e.g. when they occur between nearest neighbors \cite{Bel98,Enr99}.
The behavior difference between these two limiting regimes 
raises the question of whether there exists a critical value for the 
range of external exchanges for patterning to occur.  
The main objective of this Letter is to  address this question,
which besides having its own theoretical interest, is relevant to
alloys under irradiation.
Indeed, forced relocation of atoms in displacement cascades may extend 
beyond nearest-neighbor distances, especially in the case of dense cascades 
or for open crystal structures \cite{Ave98}.               

To render the problem more concrete, let us consider a binary alloy 
with a positive heat of mixing, under irradiation.
Each time an external
particle collides with the solid, a local atomic rearrangement is produced.
These rearrangements have a ballistic component that mixes the atoms
regardless of their chemical identity, trying to bring the system
to a random solid solution.  
Due to its local nature, the ballistic mixing will relocate atoms in a 
region of characteristic radius $R$. 
The case  $R \to \infty$, or 
{\it arbitrary-length ballistic exchanges}, has already been studied:
The macroscopic governing equation is identical to that 
describing a binary alloy undergoing a chemical reaction 
$A \rightleftharpoons B$ \cite{Glo95},
and the one describing a block copolymer (BCP) melt \cite{Liu89}.  
From the studies of these systems,
the physics of this case is well understood.  
In terms of the frequency of forced exchanges $\Gamma$,
it has been shown that while high values bring the system to a 
random solid solution, there is a critical value
below which the homogeneous concentration 
profile becomes unstable towards phase separation.  As
in spinodal decomposition, enriched regions form and coarsen.
However, the characteristic length of the domains $l$,
instead of growing indefinitely towards a macroscopic phase separation,
saturates at a mesoscopic scale, $l_\infty$.  
For $\Gamma$ values close
to the critical value, the steady-state concentration profile has a 
sinusoidal-wave appearance, with diffuse interfaces, what is 
referred to as the weak-segregation regime. For smaller $\Gamma$ values, 
the concentration profile presents sharper interfaces,
with a square-wave-like appearance, what is referred to as 
the strong-segregation regime.  
In each regime, the characteristic length has been shown 
to follow a power law with the exchange frequency 
$l_\infty \sim \Gamma^{-\phi}$, with an exponent $\phi$ of 1/4 for the 
weak
and 1/3 for the strong-segregation regime \cite{Liu89}.

In the case of ballistic exchanges of a finite range $R$, there should still
be a critical value $\Gamma(R)$ below which the system phase separates.
The question is whether phase coexistence takes place at a 
macroscopic or mesoscopic scale.
In principle, we can predict that if certain $\Gamma$ equilibrates a
wavelength $l$ for arbitrary length exchanges, finite range exchanges
must also generate patterns when $R \gg l$. It is, however, 
difficult to determine a priori the exact conditions under
which coarsening will saturate as a function of $R$ and $\Gamma$.
In this Letter we show that given a certain $R$,
there is an interval $[\Gamma_1(R),\Gamma_2(R)]$ for the stabilization of
patterns. Above $\Gamma_2$ the homogeneous concentration profile is 
stable, and below $\Gamma_1$ coarsening continues with the time,
with the system separating into macroscopic phases.  The extent
of this interval for patterning
decreases with $R$, reaching a zero value at a
critical, nonzero value $R_c$, 
when $\Gamma_1(R_c) = \Gamma_2(R_c) = \Gamma_c$.  For 
ballistic mixing with a radius smaller than $R_c$, patterning 
is not possible.  We also show that, given a mixing distance $R$, 
there is an upper bound for the wavelengths attainable as 
$\Gamma \to \Gamma_1^+(R)$.  These conclusions are in agreement with
recent Kinetic Monte Carlo simulations of binary alloys under 
finite-range ballistic exchanges \cite{Hai,Enr00}.


We study the problem using a Cahn-Hilliard-type description
of one-dimensional fronts, simulating the walls of the labyrinthine patterns,
and we construct a variational formulation to investigate the solution.
The equation describing the temporal evolution 
is composed of two terms, one for thermal diffusion and another 
one for ballistic mixing \cite{Mar84}:
\begin{equation}
\frac{\partial \psi}{\partial t} = 
\frac{\partial \psi^{\text{th}}  }{\partial t} + 
\frac{\partial \psi^{\text{bal}} }{\partial t}.
\end{equation}
Here we have chosen to represent the concentration field by a
globally conserved order parameter $\psi({\bf x})$ 
so that the homogeneous concentration
profile (solid solution) corresponds to $\psi = 0$. In the previous equation, 
the first term is simply given by 
$M \nabla^2 ( \frac{\delta F}{\delta \psi})$, where 
$F$ is the global free energy,  
and we have assumed a constant mobility $M$.
As for the second term, we actually need 
to perform a derivation.  For that purpose, let us consider first 
ballistic mixing occurring one dimensionally between planes along a 
crystallographic direction. The rate of change of concentration in the plane 
$i$ due to interchange of atoms with the planes labeled $j$ is
given by \cite{Mar84} 
\begin{equation}
\frac{ \partial \psi_i^{\text{bal}}}{\partial t} = 
- \Gamma \sum_j w_j ( \psi_i - \psi_{i+j}) = 
- \Gamma( \psi_i - \langle \psi \rangle),   
\end{equation}
where $w_j$ is a normalized weight function describing the 
distribution of ballistic exchange distances, and the brackets
denote the corresponding (discrete) weighted spatial average.
The extension to the continuum is immediate, and we write the
governing equation as:
\begin{equation}
\frac{ \partial \psi}{\partial t} = 
M \nabla^2( \frac{\delta F}{\delta \psi}) - 
\Gamma (\psi - \langle \psi \rangle_R) .
\label{eq:gov}
\end{equation} 
$w_R({\bf x})$ is now a continuous function peaked around the 
origin with a width proportional to $R$, and the average denoted by the
brackets is defined as:
\begin{equation}
\langle \psi \rangle_R = \int w_R({\bf x}-{\bf x}') \psi({\bf x}') d {\bf x}' .
\end{equation}
In the limit $R \to 0$, the ballistic
term reduces to a Laplacian term, expressing diffusion.
In limit $R \to \infty$, we recover the governing
equation for the case of arbitrary-length ballistic exchanges. 


In analogy to what was done in the case of arbitrary-length 
exchanges, we seek to find a 
Lyapunov functional for this problem, which we shall refer to as the 
free energy functional of the system \cite{note:Lya}.  
This idea actually traces back to the work of Leibler \cite{Lei80}, and 
Ohta and Kawasaki \cite{Oht86}, on block copolymer melts. 
This functional is given by $E = F + \gamma G$,
and it is built so as to determine the kinetics: 
$\frac{ \partial \psi}{ \partial t} = 
 M \nabla^2 ( \frac{\delta E}{\delta \psi})$.  
$G$ is a new term describing {\it effective interactions}
related to the ballistic term, 
and to simplify the notation,  we use $\gamma=\Gamma/M$
for the rest of this paper.

For $F$, we use a Ginzburg-Landau free energy
\begin{equation}
F = \int ( -A \psi^2 + B \psi^4 + C |\nabla\psi|^2) \ d{\bf x} ,
\end{equation}
while $G$ is expressed by a self-interaction of the form
\begin{equation}
G = \frac{1}{2} \int \int 
\psi({\bf x}) g({\bf x} - {\bf x}') \psi({\bf x}') 
\ d{\bf x} \ d{\bf x}' ,
\end{equation} 
with $g$ a kernel satisfying
\begin{equation}
\nabla^2 g({\bf x}-{\bf x}') = - (\delta({\bf x}-{\bf x}') -
                                   w_R ({\bf x}-{\bf x}') ).
\end{equation}

To proceed further, at this point we need to make a choice of
the weight function $w_R$.  A Yukawa-type potential has been proposed
by N. Goldenfeld \cite{Gol99}. This form fits
the observed distribution distances of ballistic exchanges for
crystals under irradiation \cite{Ave98}, 
while allowing us to handle part of the minimization problem analytically.
In one dimension, $w_R(u) = R/2 \exp(-|u|/R)$.  

\begin{figure}
\begin{center}
    \epsfysize=2in
    \epsfbox{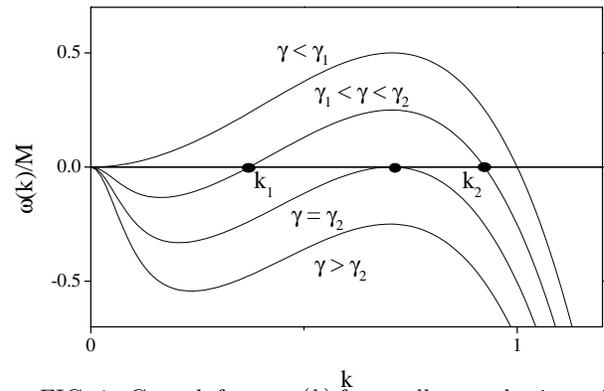}
\caption{Growth factor $\omega(k)$ for small perturbations of wave vector $k$.}
\label{fig:ampf}
\end{center}
\end{figure}

Having stated the problem, 
we start by performing a stability analysis of Eq.~\ref{eq:gov}.
For small perturbations of the
form $e^{\omega t + i k x}$ around a constant profile $\psi=0$, 
it is straightforward to find the dispersion relation:
\begin{equation}
\frac{\omega(k)}{M} = 2 A k^2 - 2 C k^4 - 
                      \frac{ \gamma R^2 k^2}{1 + R^2 k^2} .  
\end{equation}
A series of plots of this dispersion relationship for different values 
of $\gamma$ is shown in Fig.~\ref{fig:ampf}.  
As in the case of arbitrary length ballistic exchanges, 
there is a critical value $\gamma_2$ below which the
homogeneous concentration profile becomes unstable.
Below this value, there is a window of $k$ values $(k_1,k_2)$
for which the homogeneous solution is locally unstable, suggesting 
a wavelength selection.  For smaller values of 
$\gamma$, the dispersion relation resembles the one of spinodal
decomposition, suggesting 
macroscopic phase separation.

To confirm these predictions, we
use a variational approach,
based in  minimizing the free energy functional $E$.
Let us consider first the weak-segregation regime, where
the choice of $w_R$ allows us to solve the problem analytically, 
and then consider the strong-segregation regime, where we need to 
appeal to a numerical treatment.

In the weak-segregation regime, we can perform
the minimization of $E$ by considering a sine family of parametric 
functions, $\psi(x) = \alpha \sin(k x)$.  
We obtain the energy per unit length:
\begin{equation}
\langle E \rangle (\alpha,k) = 
- \alpha^2 \frac{A}{2} + \alpha^4 \frac{3 B}{8} + \alpha^2 k^2 \frac{C}{2}+
 \alpha^2 \frac{\gamma L^2}{4 (1 + k^2 L^2)} .
\end{equation}
Minimization of this energy is performed analytically, and  
concentration patterning, indicated by solutions with nonzero 
values of $k$ and $\alpha$ are found for an interval in $\gamma$.
In reality, the value of $\gamma_1$ predicted by this parametric
functions is an underestimation.
Before that value is reached, the energy
per unit length for the macroscopically separated system becomes 
lower than the energy per unit length for the sine profile.  The 
crossover point determines the actual value of $\gamma_1$ in the
weak segregation regime approximation, where patterning occurs
in the interval given by:
\begin{eqnarray}
  \gamma_1 &=& \surd {({\frac{2 \sqrt{3 C} + 
                        \sqrt{ 6 [ (3-\sqrt{6}) A R^2 -
                                  (\sqrt{6} -2) C]}}
                       {3 R^2}})} , \nonumber \\
  \gamma_2 &=& (A+C/R^2)^2/2C.
\end{eqnarray}
This interval shrinks to zero at a critical value of $R_c = \sqrt{C/A}$,
corresponding to $\gamma_c = 2 A^2/C$.
As a consequence of $\gamma_1$ being determined by a crossover
of energies, a transition towards macroscopic phase
separation occurs at a finite value of $k$.  As a result, there is
a bound in the wavelength of the patterns for a given $R$.  


In the strong-segregation regime, we need to improve over the approximation
of sine waves
for a proper evaluation of $\gamma_1$ 
away from the critical point. 
To this purpose, we propose to minimize
the free energy functional using the tanh-sine family of parametric
functions $\psi(x) =  \alpha \tanh( m/k \sin(k x))$.
The parameter $m$ serves to change the wave profile continuously from a 
sinusoidal type, to a tanh-like type with sharp interfaces, 
matching the concentration profile of an equilibrium interface.


It is easy to minimize first with respect to the parameter $\alpha$.  We 
obtain the expression, for $\alpha > 0$,
\begin{equation}
\langle E \rangle = - \frac{(A \varepsilon_1 - C \varepsilon_3 
                 + \gamma \varepsilon_4)^2}{4 B \varepsilon_2} ,  
\end{equation}
where the $\varepsilon_i$ quantities denote the energy per unit length
associated with each one of the terms in the free energy functional.
Here we can see that the values of $\gamma$ and $k$ that 
minimize the free energy are independent of the parameter $B$, 
which only relates to the amplitude $\alpha$.

\begin{figure}
\begin{center}
    \epsfysize=2in
    \epsfbox{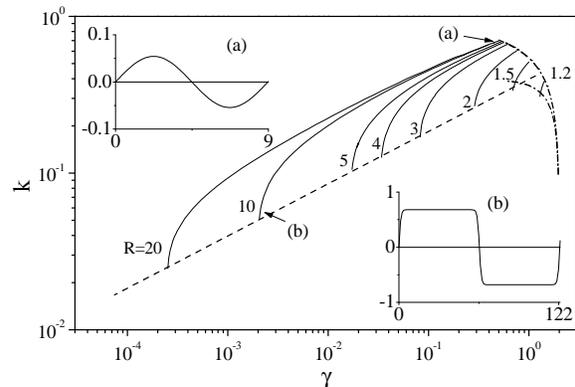}
\caption{Wave vector $k$ as a function of $\gamma$ for several values of $R$.
Dash-dotted lines mark the limits of patterning as predicted by 
the sine profiles in the weak-segregation regime.
The dashed line correspond to the large $R$ power-law fit.
Inserts (a) and (b) show the concentration profile for $R$=10 in
the weak and strong segregation regimes respectively.} 
\label{fig:tanh1}
\end{center}
\end{figure}

\begin{figure}
\begin{center}
    \epsfysize=2in
    \epsfbox{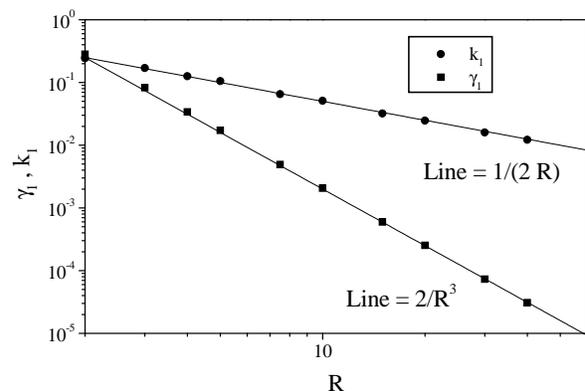}
\caption{Double logarithmic plots of $k_1$ and $\gamma_1$ versus $R$, for
large $R$, showing a power law dependency.}
\label{fig:tanh2}
\end{center}
\end{figure}

\begin{figure}
\begin{center}
    \epsfysize=2in
    \epsfbox{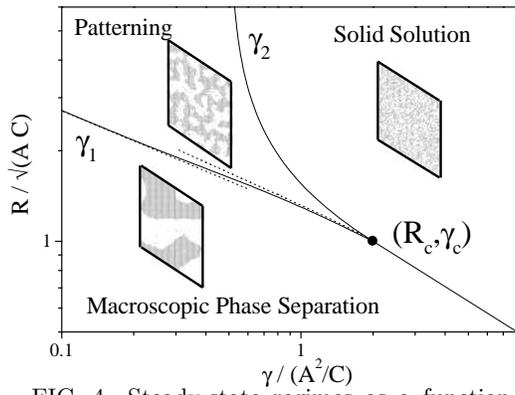}

\caption{Steady-state regimes as a function of 
$R$ and $\gamma$.  Asymptotics for $\gamma_1$ are indicated by
dashed lines, for $R \sim R_c$ (sine profiles) and 
$R \gg R_c$ (power-law fits).  Inserts show cuts of 3D Kinetic Monte 
Carlo simulations (to be presented elsewhere) 
of an FCC A$_{50}$B$_{50}$ alloy with ballistic exchanges of
range $R \sim 5 / \sqrt{2}$~lattice parameters.}
\label{fig:diag}
\end{center}
\end{figure}

The energy per unit length (the $\varepsilon$ terms)
cannot be obtained analytically for these functions, 
so we proceed with a numerical strategy.
The free energy per unit length is computed by numerical integration,
and the actual minimization is performed by means of the 
subroutine MNFB from NETLIB \cite{netlib}, 
based on a secant Hessian approximation.
Figure \ref{fig:tanh1} shows $k$ versus $\gamma$ plots for a series of 
values of $R$. The physical parameters $A$ and $C$ are set to unity.
From this plot we obtain the dependency of $\gamma_1$ and the 
corresponding wave vector $k_1$ (related to the maximum attainable wavelength)
as a function of $R$.  Figure~\ref{fig:tanh2} is a double-log plot
of these quantities for large values of $R$, showing a power law dependency.  
A fit of the data with the  power laws  $\gamma_1 = p R^{-\theta}$ 
and $k_1 = q R^{-\sigma}$ yield the quantities $p=2.26 \pm 0.03$, 
$\theta = 3.039 \pm 0.006 $ and 
$q = 0.54 \pm 0.02 $, $\sigma = 1.03 \pm 0.01 $.  
Still, an almost perfect fit is obtained by
the simple laws: $\gamma_1 = 2/R^3$ and $k_1 = 1/(2 R)$, as shown in 
the figure.  The latter relationship has the physical interpretation that
$R$ is the parameter determining the maximum wavelength.
Furthermore, the combination of these two relationships yield the
power law for the $R \rightarrow \infty$ case: $k \propto \gamma^{1/3}$. 
For values of $A$ and $C$ not equal to one, the corresponding
dependencies can be derived by dimensional analysis, giving:
$\gamma_1 = 2 \sqrt{A C}/R^3$ and $k_1 = 1/(2 R)$.
Figure~\ref{fig:diag} summarizes the steady-state 
regimes in the $\gamma$-$R$ space.
Patterning may occur in the present model 
when the range of the forced exchanges exceeds a 
critical value, with a maximum wavelength proportional to that range.
In the case of alloys under irradiation, typical $R$ values range
from 2 to 10~$\AA$, suggesting that patterns up to 100~$\AA$ could be
stabilized.  Experimental work to test these predictions is under
progress.
  

We are very grateful to Nigel Goldenfeld for his clever suggestions
and enlighting discussions. 
We also want to thank the MRL Center for Computation 
at the University of Illinois for their assistance.  
This work was partly supported by the
US Department of Energy grant DOFG02-96ER45439 through the
University of Illinois Materials Research Laboratory.

\begin{references}
\bibitem{note:Lya} The use of a Lyapunov functional as
a global stability criterion is not strictly correct, since fluctuations
are absent from Eq.~(\ref{eq:gov}). For a discussion, see Ref.~\cite{Mar97}.
\end{references}
\end{document}